\documentclass{appolb}
\usepackage{epsfig}
\usepackage{amsmath}
\usepackage{graphicx}
\usepackage{bm}

\begin{document}
\title{Glauber Monte-Carlo predictions for LHC%
\thanks{Presented at Cracow Epiphany Conference
on LHC Physics, 4 - 6 January 2008, Cracow, Poland}
}
\author{
Maciej Rybczy\'nski$^a$\footnote{E-mail: Maciej.Rybczynski@pu.kielce.pl}, Wojciech Broniowski$^{a,}$$^b$\footnote{E-mail: Wojciech.Broniowski@ifj.edu.pl}, Piotr Bo\.zek$^{b,}$$^c$\footnote{E-mail: Piotr.Bozek@ifj.edu.pl}%
\address{$^a$Institute of Physics, Jan Kochanowski University, PL-25406~Kielce, Poland\\
         $^b$The H. Niewodnicza\'nski Institute of Nuclear Physics, Polish Academy of Sciences, PL-31342 Krak\'ow, Poland\\
         $^c$Institute of Physics, Rzesz\'ow University, PL-35959 Rzesz\'ow, Poland}
}

\maketitle
\begin{abstract}
In the framework of various Glauber-like models we compute several
correlation observables in nuclear collisions at the SPS, RHIC, and  LHC energies.
We analyze fluctuations of the eccentricity of the
fireball created in the collision, in particular the
 {variable-axes} harmonic moments
$\varepsilon^\ast$, as well as  the
 fluctuations of multiplicity of charged particles.
%, in particular their scaled variances and scaled standard deviations.
We find moderate model dependence of the scaled standard deviation
$\sigma(\varepsilon^\ast)/\varepsilon^\ast$
on the choice of the particular Glauber model.
For all considered models the values of $\sigma(\varepsilon^\ast)/\varepsilon^\ast$ range from $\sim$~0.5 for central collisions to $\sim$~0.3-0.4 for
peripheral collisions.
The results are confronted to the recent measurement of the elliptic-flow fluctuations at RHIC.
We also find that the dependence of multiplicity fluctuations on the
centrality of the collision is too weak to explain the measurements at the SPS energies. The magnitude
of the Glauber multiplicity fluctuations increases by about 20\% from the RHIC to LHC energies.
\end{abstract}

\section{Introduction}

Studies of correlations and fluctutations have become a major part
of the heavy-ion-collisons physics program, as these observables may carry valuable information on the
dynamics of the system. While the dynamical nature of correlations
is of great theoretical interest, a part of the observed
effect originates from purely statistical phenomena, such as the fluctuation of the number of participants 
in a given centrality class, fluctuations of the shape and orientation of the fireball, etc.
It is thus important to understand in detail the ``background'' of these non-dynamical fluctuations.

In this talk we present predictions of a variety of Glauber-like models of the initial
stage of a heavy-ion reaction for several correlation observables studied at SPS and RHIC, which can also be measured in the forthcoming
LHC experiments.
We analyze fluctuations of the eccentricity in the transverse plane of the initial fireball, including
the {\em variable-axes} harmonic moments,
$\varepsilon^\ast$. The predictions are compared to the recent measurements of the fluctuations of the
elliptic-flow coefficient at RHIC.

New results are shown for the fluctuations of multiplicity of charged particles, and compared to the NA49 results. It is found that
the fluctuation induced by the Glauber-model are not sufficient to explain these data,
leaving room for dynamical effects.

The formalism used in this talk is described in detail in
Refs.~\cite{Broniowski:2007ft,Broniowski:2007nz}. In particular, all details
concerning the statistical methods and the variants of the Glauber models may be found there.

\section{Simulations in Glauber-like models}

With the help of a computer program
 GLISSANDO \cite{Broniowski:2007nz} we have analyzed several
 variants of Glauber-like Monte-Carlo models:

\begin{itemize}

\item The standard {\em wounded nucleon model}~\cite{Bialas:1976ed}. The wounding cross section $\sigma_{\rm w}$ is equal $32$~mb, $42$~mb, and
$65$~mb~\cite{pdg} for SPS, RHIC, and LHC energies, respectively.

\item The {\em mixed} model, amending wounded nucleons with some admixture $\alpha$ of binary collisions \cite{Back:2001xy,Back:2004dy}.
The successful fits to particle multiplicities (see Ref.~\cite{Back:2004dy}) give $\alpha = 0.145$ at $\sqrt{s_{NN}}=200$~GeV and $\alpha = 0.12$ at $\sqrt{s_{NN}}=17.3$~GeV. For LHC energy $\sqrt{s_{NN}}=5.5$~TeV we made an educated guess for the mixing parameter, $\alpha = 0.2$.

\item We also analyze a model with {\em hot spots} (see Ref.~\cite{Gyulassy:1996br}), assuming that the
cross section for a semi-hard binary collisions producing a hot-spot is tiny, $\sigma_{\rm hot-spot} = 0.5$~mb for all energies,
however when such a rare collision occurs it produces on the average a very large amount of the transverse energy
equal to $\alpha\sigma_{\rm w}/\sigma_{\rm hot-spot}$.

\item Each source from the previously described models may deposit the transverse energy with a certain
probability distribution. Thus, we superimpose the $\Gamma$ distribution, $g(w,\kappa)$, over the distribution of sources
\mbox{$g(w,\kappa)={w^{\kappa-1}\kappa^\kappa \exp(-\kappa w)}/{\Gamma(\kappa)}$.}
Here we do this superposition on the hot-spot model, labeled {\em hot-spot+$\Gamma$}.
We set $\kappa=0.5$, which gives ${\rm var}(w)=2$ for the analysis of elliptic flow fluctuations. In order to reproduce multiplicity distribution in p+p collisions at SPS energies, we set $\kappa=1$, which gives ${\rm var}(w)=1$ for analysis of multiplicity fluctuations.

\end{itemize}
The four considered models (wounded-nucleon, mixed, hot-spot, and hot-spot+$\Gamma$)
differ substantially in the number of sources and the amount of the built-in fluctuations.

\section{Event-by event fluctuations of the elliptic flow}

\begin{figure}[b]
\begin{center}
\includegraphics[width=.67\textwidth]{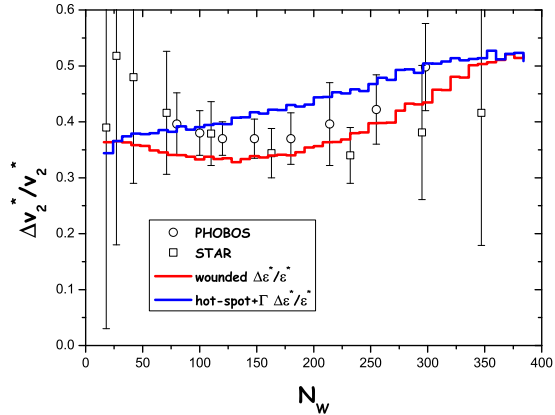}
\end{center}
\vspace{-7mm}
\caption{(Color online) Fluctuations of $\epsilon^\ast$ in two variants of the Glauber models, compared to the data from Refs.~\cite{Alver:2006wh,Sorensen:2006nw,Alver:2007rm}. See the text for details.\label{fig:v20}}
\end{figure}

The proper description of the mechanism of the
fluctuations of the elliptic flow may be  very helpful in understanding the  dynamics of heavy-ion collision, especially in its early stage~\cite{Wang:2006xz}. Moreover, for the first time
the elliptic-flow fluctuations have recently been measured at RHIC~\cite{Alver:2006wh,Sorensen:2006nw,Alver:2007rm}. The
interpretation of these results connects the fluctuations of the eccentricity coefficient, $\varepsilon^\ast$, with the
fluctuations of the variable-axes flow coefficient, denoted in this talk as $v_2^\ast$.
For sufficiently small elliptic asymmetry, which is an experimental fact, one expects on purely hydrodynamic grounds the relation
\begin{eqnarray}
\frac{\sigma( v_2^\ast)}{v_2^\ast}=\frac{\sigma( \varepsilon^\ast)}{\varepsilon^\ast}. \label{hydrofl}
\end{eqnarray}
Comparison of the data to our Glauber calculations is made in Fig.~\ref{fig:v20}. For central collisions we expect
the result following from the central limit theorem \cite{Broniowski:2007ft} for uncorrelated sources,
\begin{eqnarray}
\sigma(v_2^\ast)/v_2^\ast\simeq\sigma(\varepsilon^\ast)/\varepsilon^\ast = \sqrt{4/\pi-1} \simeq 0.52 \;\;\;\;\;(b=0), \label{v2res}
\end{eqnarray}
which is compatible with the data, although the error bars are large. We note that, with the identification
(\ref{hydrofl}), the Glauber models reproduce properly the data for not-too-peripheral collisions, where the approach is credible.
Other variants fall between the wounded-nucleon model, which has the lowest amount of fluctuation, and the hot-spot+$\Gamma$ model, which has the strongest fluctuations. More accurate measurements are needed to discriminate the model predictions. We note, however, that the statistical
fluctuations built in the Glauber model are sufficient to explain the data.

\begin{figure}[b]
\begin{center}
\includegraphics[width=.67\textwidth]{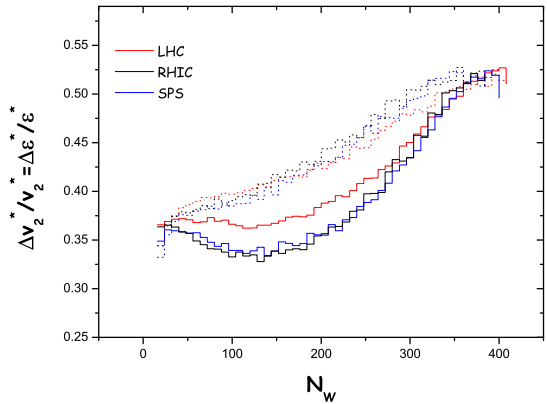}
\end{center}
\vspace{-7mm}
\caption{(Color online) Fluctuations of $v_2^\ast$ at SPS, RHIC, and LHC energies from the Glauber model with wounded nucleons only (full lines) and the hot-spot$+\Gamma$ model (dotted lines) .\label{fig:v2}}
\end{figure}

In Fig.~\ref{fig:v2} we compare the results of simulations of the elliptic-flow fluctuations at SPS and RHIC with the predictions for the LHC energies.
The calculation is done for two extreme variants of Glauber-like models, namely the wounded nucleon model, where the fluctuations of initial shape are smallest, and the model with hot-spots+$\Gamma$, where the initial shape fluctuates are most prominent.
For the most central and most peripheral collisions
the results are very similar for both models at all energies, however at intermediate centralities  there are much higher fluctuations
for the LHC energies with the hot-spots+$\Gamma$ model than in other cases, as expected.

\section{Multiplicity fluctuations}

\begin{figure}[b]
\begin{center}
\includegraphics[width=.67\textwidth]{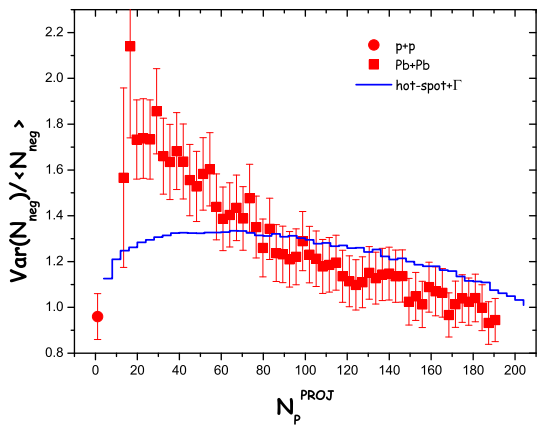}
\end{center}
\vspace{-7mm}
\caption{(Color online) Scaled variance of the multiplicity distribution of negatively charged hadrons produced in p+p and Pb+Pb collisions at
the top SPS energy, plotted as a function of centrality determined by the number of projectile participants measured by NA49 experiment. The line 
is obtained with the hot-spot$+\Gamma$ model with proper implementation of the experimental acceptance.
The data are from Ref.~\cite{Alt:2006jr}.\label{fig:na49}}
\end{figure}

Recently the NA49 Collaboration published results~\cite{Alt:2006jr}
on centrality and system-size dependence of multiplicity fluctuations observed
in Pb+Pb minimum bias and in p+p collisions.
Unexpectedly, the scaled variance ${\rm Var}(N)/\langle N \rangle$,
where ${\rm Var}(N)$ is the variance and $\langle N \rangle$  the
average multiplicity of the observed charged particles, changes
non-monotonically when the number of wounded nucleons grows. The scaled variance
is close to unity in peripheral and
central collisions, however it shows a very prominent
peak at $N_w \approx 70$. The measurement has been performed at
the top SPS collision energy $\sqrt{s_{NN}}=17.3$~GeV in the transverse momentum and
pion rapidity intervals $(0.005,1.5)$~GeV and $(1.1,2.6)$,
respectively. The azimuthal acceptance has been also limited, and only
about $20\%$ of all produced negative particles have been used in
the analysis.
In Fig.~\ref{fig:na49} we show NA49 results for the negatively-charged particles compared to 
the Glauber-like simulation of the  model  hot-spots+$\Gamma$ in the NA49 azimuthal acceptance. 
The experimental acceptance of $20\%$ has been implemented in the simulation.
Despite the fact that the fluctuations in the hot-spot+$\Gamma$ model are the highest of all
considered Glauber models, it is evident from
the plot that the Glauber calculation alone cannot describe the  non-monotonic dependence
of scaled variance on centrality. Near the peak the model falls below the data, while
at larger $N_w$ it goes above the data.
For other models the obtained curve is even lower. In our view
dynamical effects must be incorporated in order to understand the phenomenon \cite{Baym:1995cz,Rybczynski:2004zi}.

\begin{figure}[b]
\begin{center}
\includegraphics[width=.67\textwidth]{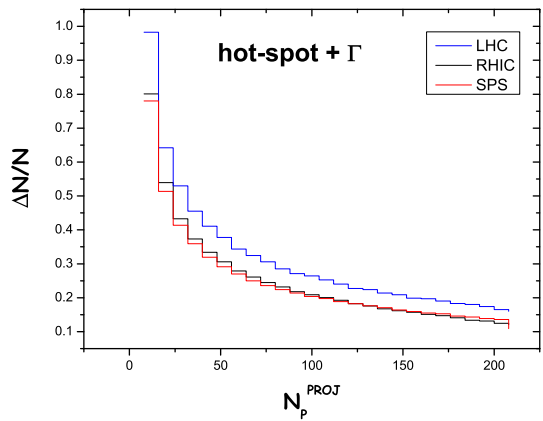}
\end{center}
\vspace{-7mm}
\caption{(Color online) Scaled standard deviation of the multiplicity distribution of negatively charged hadrons plotted as a function of the number
of projectile participants for SPS, RHIC and LHC energies, all results for the Glauber Monte-Carlo hot-spot$+\Gamma$ model. \label{fig:lhc}}
\end{figure}

In Fig.~\ref{fig:lhc} we compare the centrality dependence of the model multiplicity
fluctuations for the SPS, RHIC, and LHC energies in the full acceptance. Here we use scaled standard deviation of multiplicity distribution as a measure of multiplicity fluctuations.
Multiplicity fluctuations at SPS and RHIC energies are very close to each other, however we notice an increase of
for the LHC by about $20\%$. This effect is due to the larger value of the wounding cross section $\sigma_w$.
The conclusion is qualitative, since as mentioned above, the Glauber models do not explain
the details of the SPS measurement of the multiplicity fluctuations.

\section{Conclusions}

Our main results are as follows:

\begin{itemize}

\item The fluctuations of the eccentricity of the fireball are
related to the fluctuations of the elliptic flow
coefficient. They depend rather moderately on the chosen Glauber model. Our results agree with the
recent PHOBOS and STAR measurements of the $v_2$ fluctuations. The agreement indicates
that this quantity is dominated by the Glauber-model statistics.

\item The Glauber-model predictions for the multiplicity fluctuations fail short of the experimental results of Ref.~\cite{Alt:2006jr},
leaving room for strictly dynamical effects \cite{Baym:1995cz,Rybczynski:2004zi}. Such effects should
be introduced in order to understand the non-monotonic
dependence of the scaled variance of multiplicity on the number of produced particles.

\end{itemize}

\bigskip

\bigskip

This research has been supported by the Polish Ministry of Science and Higher Education under grants N202~034~32/0918
and 1~P03B~127~30.


\begin{thebibliography}{24}
\expandafter\ifx\csname natexlab\endcsname\relax\def\natexlab#1{#1}\fi
\expandafter\ifx\csname bibnamefont\endcsname\relax
  \def\bibnamefont#1{#1}\fi
\expandafter\ifx\csname bibfnamefont\endcsname\relax
  \def\bibfnamefont#1{#1}\fi
\expandafter\ifx\csname citenamefont\endcsname\relax
  \def\citenamefont#1{#1}\fi
\expandafter\ifx\csname url\endcsname\relax
  \def\url#1{\texttt{#1}}\fi
\expandafter\ifx\csname urlprefix\endcsname\relax\def\urlprefix{URL }\fi
\providecommand{\bibinfo}[2]{#2}
\providecommand{\eprint}[2][]{\url{#2}}

%\cite{Broniowski:2007ft}
\bibitem{Broniowski:2007ft}
  W.~Broniowski, P.~Bo\.zek and M.~Rybczy\'nski,
  %``Fluctuating initial conditions in heavy-ion collisions from the Glauber
  %approach,''
  Phys.\ Rev.\  C {\bf 76}, 054905 (2007), \eprint{0706.4266 [nucl-th]}.
  %%CITATION = PHRVA,C76,054905;%%

\bibitem{Broniowski:2007nz}
\bibinfo{author}{\bibfnamefont{W.}~\bibnamefont{Broniowski}}
  \bibinfo{author}{\bibfnamefont{M.}~\bibnamefont{Rybczy\'nski}}, \bibnamefont{and}
  \bibinfo{author}{\bibfnamefont{P.}~\bibnamefont{Bo\.zek}}
  (\bibinfo{year}{2007}), \eprint{arXiv:0710.5731 [nucl-th]}.

\bibitem{Bialas:1976ed}
\bibinfo{author}{\bibfnamefont{A.}~\bibnamefont{Bia\l{}as}},
  \bibinfo{author}{\bibfnamefont{M.}~\bibnamefont{B\l{}eszy\'nski}},
  \bibnamefont{and} \bibinfo{author}{\bibfnamefont{W.}~\bibnamefont{Czy\.z}},
  \bibinfo{journal}{Nucl. Phys.} \textbf{\bibinfo{volume}{B111}},
  \bibinfo{pages}{461} (\bibinfo{year}{1976}).

\bibitem{pdg}
\bibinfo{author}{\bibfnamefont{W.-M.} \bibnamefont{Yao}} \bibnamefont{et~al.}
  (\bibinfo{collaboration}{Particle Data Group}), \bibinfo{journal}{J. Phys.}
  \textbf{\bibinfo{volume}{G33}}, \bibinfo{pages}{1}
  (\bibinfo{year}{2006}).

\bibitem{Back:2001xy}
\bibinfo{author}{\bibfnamefont{B.~B.} \bibnamefont{Back}} \bibnamefont{et~al.}
  (\bibinfo{collaboration}{PHOBOS}), \bibinfo{journal}{Phys. Rev.}
  \textbf{\bibinfo{volume}{C65}}, \bibinfo{pages}{031901}
  (\bibinfo{year}{2002}).

\bibitem{Back:2004dy}
\bibinfo{author}{\bibfnamefont{B.~B.} \bibnamefont{Back}} \bibnamefont{et~al.}
  (\bibinfo{collaboration}{PHOBOS}), \bibinfo{journal}{Phys. Rev.}
  \textbf{\bibinfo{volume}{C70}}, \bibinfo{pages}{021902}
  (\bibinfo{year}{2004}).

\bibitem{Gyulassy:1996br}
\bibinfo{author}{\bibfnamefont{M.}~\bibnamefont{Gyulassy}},
  \bibinfo{author}{\bibfnamefont{D.~H.} \bibnamefont{Rischke}},
  \bibnamefont{and} \bibinfo{author}{\bibfnamefont{B.}~\bibnamefont{Zhang}},
  \bibinfo{journal}{Nucl. Phys.} \textbf{\bibinfo{volume}{A613}},
  \bibinfo{pages}{397} (\bibinfo{year}{1997}).

\bibitem{Wang:2006xz}
\bibinfo{author}{\bibfnamefont{G.}~\bibnamefont{Wang}},
  \bibinfo{author}{\bibfnamefont{D.} \bibnamefont{Keane}},
  \bibinfo{author}{\bibfnamefont{A.}~\bibnamefont{Tang}},
  \bibnamefont{and}
  \bibinfo{author}{\bibfnamefont{S.~A.}~\bibnamefont{Voloshin}},
  \bibinfo{journal}{Phys. Rev.} \textbf{\bibinfo{volume}{C76}},
  \bibinfo{pages}{024907} (\bibinfo{year}{2007}).

\bibitem{Alver:2006wh}
\bibinfo{author}{\bibfnamefont{B.}~\bibnamefont{Alver}} \bibnamefont{et~al.}
  (\bibinfo{collaboration}{PHOBOS}) (\bibinfo{year}{2006}),
  \eprint{nucl-ex/0610037}.

\bibitem{Sorensen:2006nw}
\bibinfo{author}{\bibfnamefont{P.}~\bibnamefont{Sorensen}}
  (\bibinfo{collaboration}{STAR}) (\bibinfo{year}{2006}),
  \eprint{nucl-ex/0612021}.

\bibitem{Alver:2007rm}
\bibinfo{author}{\bibfnamefont{B.}~\bibnamefont{Alver}} \bibnamefont{et~al.}
  (\bibinfo{collaboration}{PHOBOS}) (\bibinfo{year}{2007}),
  \eprint{nucl-ex/0701049}.

\bibitem{Alt:2006jr}
\bibinfo{author}{\bibfnamefont{C.}~\bibnamefont{Alt}} \bibnamefont{et~al.},
(\bibinfo{collaboration}{NA49}),
  \bibinfo{journal}{Phys. Rev.}
  \textbf{\bibinfo{volume}{C75}}, \bibinfo{pages}{064904} (\bibinfo{year}{2007}),
  \eprint{nucl-ex/0612010}.

%\cite{Baym:1995cz}
\bibitem{Baym:1995cz}
  G.~Baym, B.~Blattel, L.~L.~Frankfurt, H.~Heiselberg and M.~Strikman,
  %``Correlations and fluctuations in high-energy nuclear collisions,''
  Phys.\ Rev.\  C {\bf 52}, 1604 (1995), \eprint{nucl-th/9502038}.
  %%CITATION = PHRVA,C52,1604;%%

%\cite{Rybczynski:2004zi}
\bibitem{Rybczynski:2004zi}
  M.~Rybczy\'nski and Z.~W\l{}odarczyk,
  %``Non-monotonic behavior of multiplicity fluctuations,''
  J.\ Phys.\ Conf.\ Ser.\  {\bf 5}, 238 (2005), \eprint{nucl-th/0408023}
  %%CITATION = 00462,5,238;%%

\end{thebibliography}
\end{document}